# Tuning the phase in asymmetric multilayers

Giuseppina Simone

Phase modulation has come to be recognized as a fundamental paradigm for optical device design in applications involving the spatiotemporal control of optical wavefronts. Here, asymmetric resonant multilayer cavities based on the inclusion of absorptive and lossless layers combined with the ideal conductive metal allow for reflection and phase shift control in analogy with the Fabry-Pérot system, which in its basic configuration has zero reflection and perfect transmission., the reflectivity of multilayers activated in a prism-coupled geometry and composed of numerous stack layers but built in different sequences has been examined. The layer's different arrangements, resulting in open and closed asymmetric multilayers, display reflectance spectra with distinct features as well as phase changes. For one of the studied multilayer arrangements, two hybridized resonance modes form a strong coupling with the exciton available from the red dye. As a result, energy-based topological considerations identify a correlation between exciton/polariton strong coupling, topological darkness, and phase tuning and they motivate the $3\pi$ range.

## Introduction

Recently, a variety of techniques have been established for improving phase control[1], boosted by the demonstration that its transduction guarantees an important impact on sensing [2]. A large phase shift confers to the optical systems a high flexibility that is achieved when the ratio $|\Delta\omega/\Lambda|$ between the frequency shift $\Delta\omega$ and the spectral width $\Lambda$ are maximized [3]. Accounting for the resonance frequency $\omega_0$ of the electromagnetic mode and the electric field related to the resonance $E(r)$, $\Delta\omega = -\frac{\omega_0}{2}\frac{\int dv\, \Delta\varepsilon(r)|E(r)^2|}{\int dv\, \varepsilon(r)|E(r)^2|}$ [4] and $\Lambda = 2(\gamma_r + \Gamma)$, where one can note that $\Gamma = \frac{\int dv\, \text{Im}(\varepsilon)|E(r)^2|}{\int dv\, \varepsilon|E(r)^2|}$ is the dissipative loss rate (e.g. losses ($\text{Im}(\varepsilon)$) versus the total energy) and $\gamma_r$ is the radiative rate.

The ratio $|\Delta\omega/\Lambda|$ is maximized if $\frac{\frac{\omega_0}{2}\frac{\int dv\, \Delta\varepsilon(r)|E(r)^2|}{\int dv\, \varepsilon(r)|E(r)^2|}}{2(\gamma_r + \frac{\int dv\, \text{Im}(\varepsilon)|E(r)^2|}{\int dv\, \varepsilon|E(r)^2|})} \leq \frac{C \max(r)\,|\Delta\varepsilon(r)|}{\min(r)|Im[\varepsilon(r)]|}$ and depends on the material permittivity precluding its applicability to all systems and materials with behavior not identifiable with a Drude model [5,6]. Metasurfaces have a high extent of tunability and can vary the local phase and intensity of reflected light according to the refractive index of the constituent materials [7,8]. However, the 2π phase tuning obtained by doubling the usual 0-π of the single resonance is pursued when radiated light from the electrons having a delayed phase is not absorbed [9] and, therefore, the limited variations in the metasurface optical indices [10,11] make it difficult to pursue complete tunability across the full 2π phase range while keeping a stable and large light amplitude. The simultaneous change in the amplitude and phase of the light that occurs from altering the resonance of the atoms, affecting both the phase and the intensity of the reflected light, is crucial. To meet this need, the metasurface design must have a narrow spectral width, for achieving a radiative rate to be increased against the dissipative loss rate $\gamma_r \gg \Gamma$, and a maximized resonance frequency shift [11] guaranteeing a strong coupling between the resonant mode of the metasurface and the incoming light.

Due to the behavioral diversity and independence of the pricey nanofabrication techniques required for producing metamaterials, Fabry-Pérot multilayer architectures have replaced cleanroom-made systems and found numerous optical applications associated with the improvement of the plasmonic performance derived from the phase modulation [12–15]. The fundamental configuration relying on an intermediate dielectric layer in between two perfectly reflective mirrors theoretically has zero reflection and enables a wide phase tuning of the Fabry-Pérot systems by a perfect transmission at a resonance wavelength, depending on the dielectric layer thickness [16,17]. Here, asymmetric resonant multilayer with absorptive metal, lossless dielectric, and an ideal conductive metal are strategically made [18] [19]. Silver Ag, chosen as the ideal conductive material, indium tin oxide ITO and glass $SiO_2$, the absorptive metal, and the lossless dielectric layers are all included in the layouts. ITO regulates the multilayer's reflection, while the intermediate lossless dielectric layer modifies the structure's overall phase shifts and the resonance wavelength [20]. Rhodamine has been adsorbed on top of the suggested layouts to serve as an absorbent dielectric layer. The reflectivity of opportunely arranged multilayers is activated in a prism-coupled geometry. Numerical and experimental investigations are proposed for characterizing the phase modulation and the energy in relation to the architecture. Energy and topological considerations highlight a phase tuning greater than 2π in multilayers with a strong exciton-cavity mode coupling and enable an exceptional point identified as the topological condition of the strong coupling.

## Results

**Plasmon polaritons and exciton in multilayer; mode hybridization and coupling**

The multilayers were produced using the ideal conductor Ag, ITO, an absorptive metal, glass SiO$_2$, and ((3-mercaptopropyl) trimethoxysilane), the lossless dielectrics, the latter of which was used to space the Ag and ITO or SiO$_2$ according to the method outlined in the Experimental section (see also Supporting Information, Part I). I achieved varied sample layouts by varying the layer sequence (Fig. 1a).

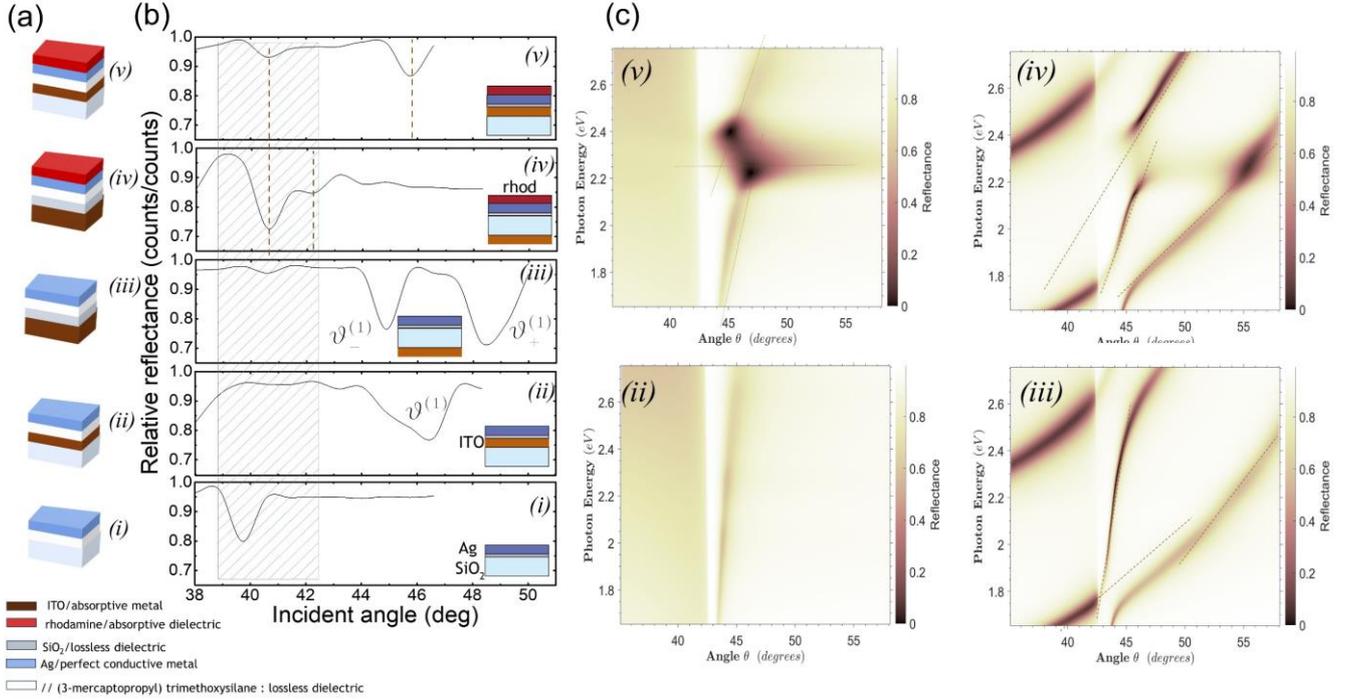

**Fig.1. Reflection from surface plasmon resonance. (a) Out-of-scale multilayer layouts. Legenda: color-coded layer. (b) Relative reflectance at TM and 650 nm. (c) Numerical angular dispersion; color bar: reflectance, insets dashed lines follow the modes.**

According to a method previously established [21] [22], a spectrophotometer was used to record the reflectance (Fig.1b), and the signal intensity was normalized by a reference to account for the uneven illumination [23]. In (ii) (Fig.1a), Ag is beneath ITO and makes an asymmetric multilayer with a reflectance spectrum displayed in Fig.1b(ii). The material losses justify the large linewidth dip at $\vartheta^{(1)} = 46.4$ deg, while prospecting a low-quality factor. In contrast, the layout (iii) with the ITO at the bottom, which acts as a second mirror (Fig.1a(iii)), has a reflection spectrum characterized by having two dips displaced to high and low angles, $\vartheta^{(1)}_+$ and $\vartheta^{(1)}_-$ (dashed lines in Fig.1b). A comparison with Ag reflectivity, used as a reference (Figs. 1a,b(i)) underlines that the dip at $\vartheta = 39.8$ deg induced by the silver surface plasmon polaritons is not apparent in the spectrum shown in figure 1b(iii), but it is moved toward a higher incident angle by the surface plasmon polaritons excited by the inner layer in Fig. 1(ii). Additional information about the modes of the multilayers (ii) and (iii) derives from the reflectance angular dispersion (Fig. 1c). A linewidth widening caused by an increase of the losses recorded as the photon energy raises characterizes the dispersion of the multilayer (ii) (Fig.2(ii))[24]; while, in the case of (iii), the dispersion shows the mode hybridization and coupling between a broad resonance mode having a large angular shift and a spectrally narrow resonance mode, which can lead to a small mode volume (Fig. 1c(iii)). Indeed, by combining the effective volume $V_{eff}$ from geometrical considerations ($V_{eff} = 5 \times 10^{-16} m^3$) and the diffraction volume $V_0 = 3\frac{(c/f)^3}{4\pi^2}$ ($= 7.14 \times 10^{-21} m^3$; $f$ is approximated by frequency @ 633 nm in air, while c is the speed of light in SiO$_2$, $c = 2 \times 10^8\ m\ s^{-1}$), the volume mode factor $V = \frac{V_{eff}}{V_0}$ can be estimated as $V = 7 \times 10^4$.

For the topological considerations that will be established in this study, it is interesting to classify the multilayer (ii) as a closed boundary system based on the reflectivity and dispersion, whereas the multilayer (iii) is hybrid because it incorporates both open and closed boundary features. Rhodamine was then adsorbed on (ii) and (iii) and the reflectance of the two new configurations (iv) and (v) was measured. The fluorescent dye acts as a molecular spring-mass oscillator and connects to one end of the cavity while a pump laser causes the molecules to transition from a ground $|g\rangle$ to an excited $|e\rangle$ state [25]. With reference to the multilayer (iii) modified into the configuration (iv) after the adsorption of the rhodamine, one observes that at an emitting frequency $\omega_m$, the dipole shows an intrinsic radiative decay rate $\gamma_m$ that can influence the multilayer and a negligible non-radiative decay rate; the latter is indeed not taken into consideration here. The multilayer in turn has a plasmonic loss $\gamma_{spp}$ at frequency $\omega_{spp}$ and, because of the crystalline Ag, it loses a part of its energy to heat at a rate $\gamma_{ohm}$. Through the molecular dipole and geometrical factors, the decay rate modifies as $\kappa^* = \frac{V_{eff}}{V_0}\frac{\gamma_m^2}{\gamma_{spp}}$ [26], which in the nearfield assumes a value $\kappa^* = 1.9\ ps^{-1}$ corresponding to a quality factor $Q = \frac{\omega_{spp}}{\kappa^*}$ ($= 1.1 \times 10^9$) (Table 1). Besides, according to the equation $\gamma'_m = \gamma_m\ Q/V$, the mode volume and the quality factor modify the radiative rate of the molecular dipole and in the updated layout, the multilayer/dye has a nearfield radiation rate schematized as $\gamma_f = \gamma'_m \frac{\kappa^*}{\kappa^* + \gamma_{ohm}}$ and, altogether, one can conclude that the radiative rate outperforms the cavity decay of a factor $\frac{\gamma_f}{\kappa^*} = 2.3$. The estimation of the ratio $\frac{\gamma_f}{\kappa^*}$ put in evidence the increment of the losses following the adsorption of the Rhodamine.

**Table 1. Factors and values of the system**

| Variable | $\gamma_m$[27] | $\omega_m$ | $V$ | $\gamma_f$ | $\gamma_{ohm}$ | $\gamma_{spp}$ | $\kappa *$ | $Q$ | $\gamma'_m$ | $\omega_{spp}$ |
|---|---|---|---|---|---|---|---|---|---|---|
| Value | $285\ ps^{-1}$ | $5.5 \times 10^8\ MHz$ | $7 \times 10^4$ | $4.5\ ps^{-1}$ | $2 \times 10^6 ps^{-1}$ | $3 \times 10^9\ ps^{-1}$ | $1.9\ ps^{-1}$ | $1.1 \times 10^9$ | $4.7 \times 10^6 ps^{-1}$ | $2.2 \times 10^9\ MHz$ |

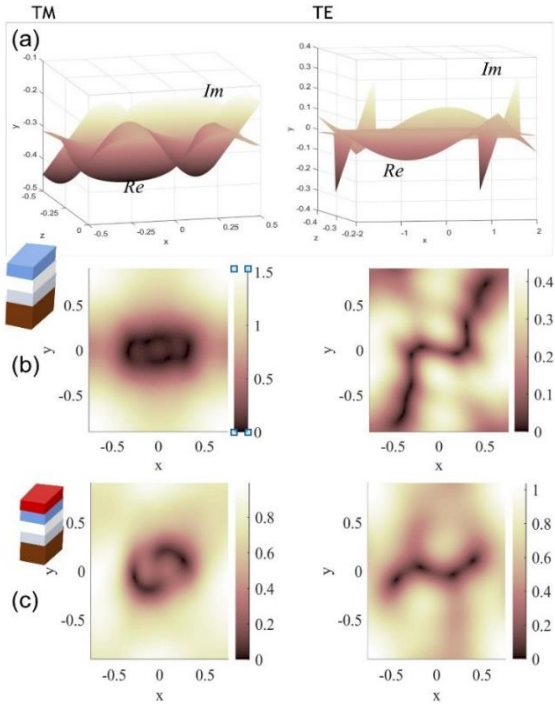

**Fig. 2. Energy split in strong coupling. (a) Real and imaginary plot of the evanescent electric field at TM and TE polarization. The cross-section in the x-y plane provides the electric field at the interface Ag/air. (b,c) Electric field energy density ($\frac{E^2 \varepsilon_0}{2}$) of (iii) (b) and (iv) (c). Color bar: $\frac{|E^2|}{|E_0^2|}$.**

There is a minimum of the reflectance at $\vartheta = 41.5$ deg combined with a small dip at $\vartheta = 42.5$ deg in the (iv) spectrum ((iv) in Fig. 1a, and Fig. 1b). One of (iii)'s two modes broadens because of the loss energy augment; in turn, the increase of energy triggers a diminution of the frequency shift and a rise of the mode volume according to the dispersion (Fig.1c(iv))[28]. In addition, the spectrally narrow mode is split by the effect of the strong coupling with the dye at a photon energy between 2.2-2.4 eV. The reflected signal trace of (v) (Fig.1a,b(v)) displays a minimum at $\vartheta = 46$ deg along with a second dip at $\vartheta = 41.5$ deg. A back-bending of the reflectance dispersion emerges from the plot in Fig. 1c(v), which proves the presence of the loss insinuated by the dye into (ii), and a weak coupling between the exciton and the polaritons. It follows that the conceived sequence of the layers confers to the polariton/exciton system a degree of freedom to control the coupling strength. In agreement with the result, the stacking order regulates the layer interference and influences the reflectivity with an impact more important than the dielectric constants[29].

To further characterize the exciton/polariton coupling in the multilayer (iv), I have accounted for the evanescent electric field energy as a measurement of the reflectance energy.

At the interface with the surrounding medium, the evanescent electric field is described by equations including an imaginary and a real part[30] according to the following formalism $E(r) = Re(E(r)) + i\,Im(E(r))$, which becomes

at TM polarization

$$E_x = \frac{2\cos\vartheta\,\sqrt{(\sin^2\vartheta - n^2)}}{(n^4\cos^2\vartheta + \sin^2\vartheta - n^2)} E^i \exp\left(-i\left(\tan^{-1}\left(\frac{\sqrt{(\sin^2\vartheta - n^2)}}{n\cos\vartheta}\right) + \frac{\pi}{2}\right)\right);$$

$$E_z = \frac{2\cos\vartheta\,\sin^2\vartheta}{(n^4\cos^2\vartheta + \sin^2\vartheta - n^2)} E^i \exp\left(-i\tan^{-1}\left(\frac{\sqrt{(\sin^2\vartheta - n^2)}}{n\cos\vartheta}\right)\right);$$

at TE polarization

$$E_y = \frac{2\cos\vartheta}{\sqrt{(1-n^2)}} E^i \exp\left(-i\frac{\frac{\left(\sin^2\vartheta - n^2\right)^2}{\cos\vartheta}}{2}\right);$$

The plots of the surface areas of the multilayer (iii) are shown in Fig. 2a while their projections in the *xy* plane are displayed in Fig. 2b. The intersection gives origin to a closed loop at TM, while it leads to an open projection, which extends to infinity at TE. Upon including an additional surface, the complex plane, which represents the exciton of the rhodamine perturbs the surface arrangement; a TM, the intersection creates a gap in space and swirls around the polarized points along x, while at TE, the intersection creates a swirl in the plane but is not sufficiently perturbative for generating a gap (Fig. 2c)[31,32]. The energy split recorded for the multilayer (iv) finds correspondence in the reflectivity energy distribution shown in the momentum space (Fig. 3). For the sake of completeness the is independent of the exciton influence when the dye is adsorbed (v).

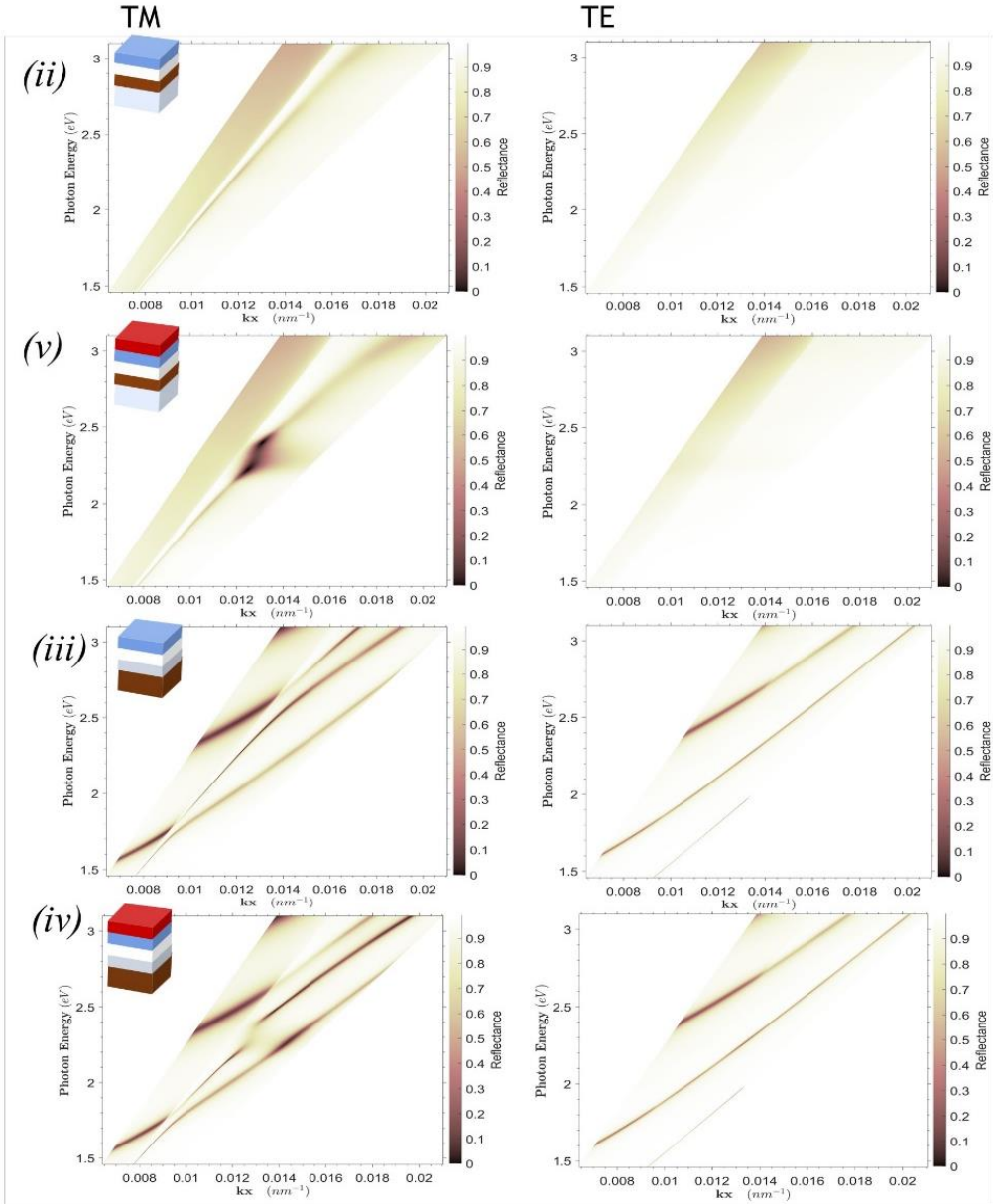

**Fig. 3. Numerical photon energy versus $k_x$ at TM, left column and at TE, right column.**

results relative to the four multilayers are reported, at transverse magnetic TM and transverse electric TE polarizations and according to $k_x$. The anti-crossing energy displayed by (iii) proves the mode hybridization, while exciton/polariton strong coupling developed in (iv) causes the energy mode split. However, again, at TE, the anti-crossing disappears, and the energy is invariant with the exciton. The multilayer (ii) and (v) do not show anti-crossing coherently with the results shown in Fig.1. At TM, a weak signal is shown by (ii), while, the energy loss rises with dye adsorption, and the energy spectrum indeed shows evidence of a weak exciton/polariton mode interaction and back-bending (dark brown region). Yet, at TE, the spectrum

## Characterizing the phase and reflectivity amplitude relying on topological considerations

The arrangement of the multilayers, designed to assemble the ITO beneath the ideal conductive metal or the lossless dielectric, results in framing into an infinite domain of both the semi-open cavity (iii) and the closed cavity (ii)[33]; in turn, both layouts are modified into semi-open configurations by the absorptive rhodamine [34]. The unicity of (iii) relies on the hybridization of the mode, and the subsequent anti-crossing and energy splitting of the spectrally narrow mode following the adsorption of the exciton (iv). However, the analysis of the reflectivity made by the spectrophotometer provides a characterization of the strong coupling, but there is no direct relationship with the phase. To this purpose, the multilayer reflectivity was measured in the optomechanical arrangement shown in Fig. 4a where the reflectivity was recorded as amplitude and phase of the electrical signal (Fig.S2, Supporting information, Part II). In this instance, the amplitude of the signal was described as $r = Re(r) + i\, Im(r)$, with $Im(r)$ and $Re(r)$ related to the dissipative loss rate and radiative rate [35][36]. The strength of the coupling assessed by the hierarchy $\gamma_r \gg \Gamma$ between the resonant mode and the incident light can be formulated as $Re(r) \gg Im(r)$, which results in the estimation of the strength through the evolution of $r$ in the complex plane. The signal's amplitude $r$ and phase shift $\varphi$ were investigate with reference to the four samples. According to the results shown in Fig. 4, a uniform amplitude is read from the layout (ii), and $Re(r) \approx Im(r)$ (Fig.4b). In the angular framework, a $\sim 2\pi$ phase modulation was measured (Fig. 4c). Even though the behavior maintains the typical characteristics of (ii) and a uniform amplitude distribution was observed, the layout (v) reads a contraction of loop (Fig. 4d)[37]. The contraction measures a drop in the coupling strength between the light source and the polaritons along with an increase of dissipating energy triggered by the interaction between the multilayer and the dye. The phase modulation spreads over a phase range < $2\pi$ (Fig.4c). The shift of (v) with respect to (ii) is shown in the angular plot of the apparent power ($\tan(arg(r))$ (Fig. 4e)), where one can also note that the power minima coincide with the phase inflections. The amplitude's topology of (iii) deviates from being a perfect circle meaning that radius of the phasor $r$ decreases because of the losses in the $\omega t$ domain (Fig. 4f) (Fig.S3, Supporting information, Part III); besides, the bulge origins between the two minima of $\tan(arg(r))$ (Fig. 4g); the phase modulation is measured (Fig. 4c). In contrast to layouts (iii) and (ii), the average rate of $\tan(arg(r))$ angular propagation decreases while approaching high incident angles (dashed lines in Fig. 4). Nevertheless, depending on whether the dye is adsorbed on (ii) or (iii), a phase modulation of about $2\pi$ or $3\pi$ is observed; therefore, when utilizing the energy from the averted crossover, a resonance shift and the capacity to walk a $3\pi$ phase modulation is granted by the multilayer (iv). This insight yields a criterion that describes the phase modulation of the Fabry-Pérot multilayer without regard to the permittivity.

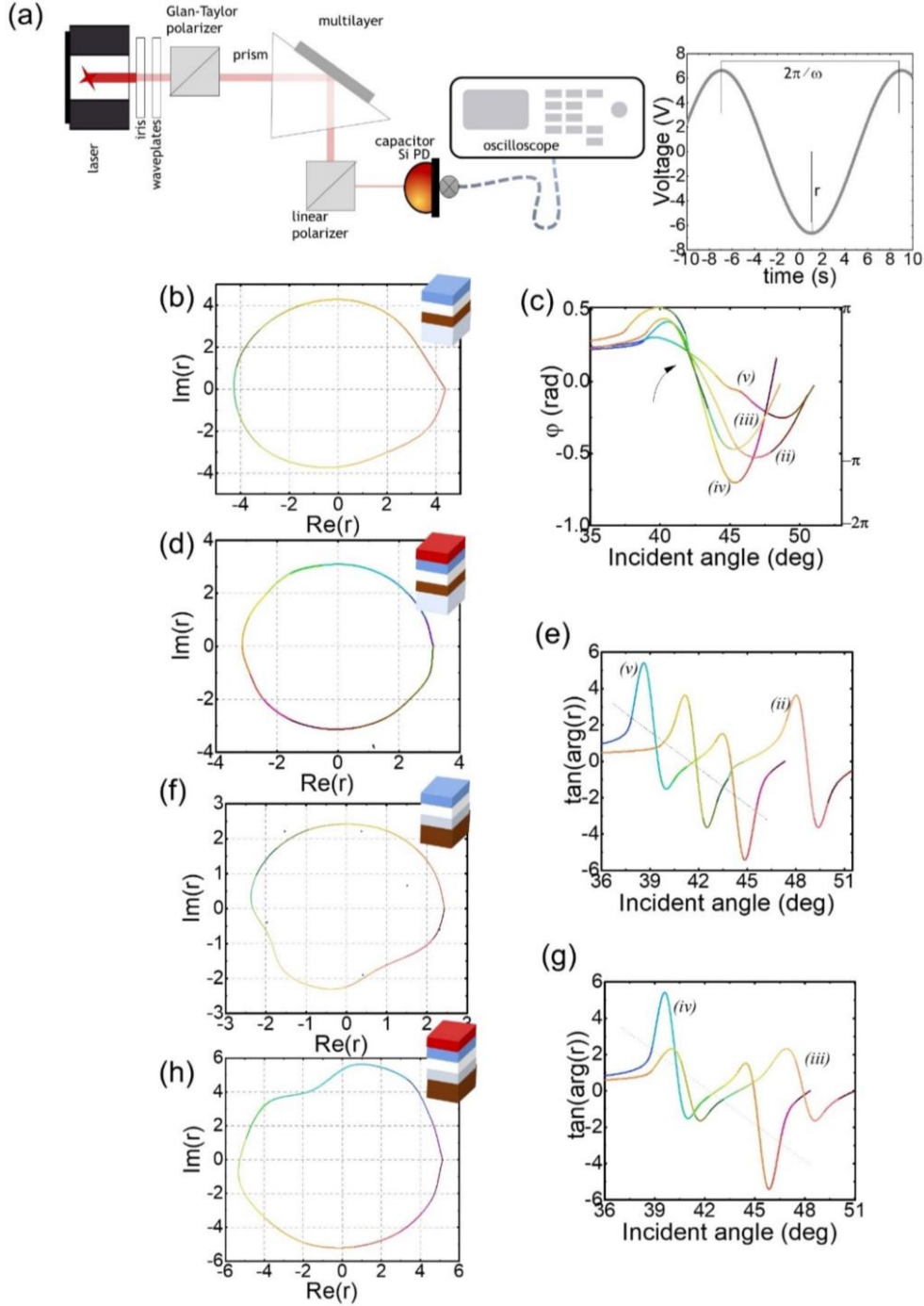

**Fig. 4. Resonance amplitude and phase modulation. (a)** Setup and signal schematic. **(b,d,f,h)** Dependence of the experimental $Im(r)$ and $Re(r)$ in the complex plane (insets: layout). **(c)** Measured angular resolved phase $\varphi$ and right axis $[-2\pi\ \pi]$ phase modulation; arrow pointing to the inflection. **(e,g)** Angular resolved $\tan(arg(r))$; dashed line averages *(iv)* and *(v)* angular trend. Color-coded plot: amplitude at the coded signal frequency

## Discussion

Here, I discuss the relationship between the phase and coupling intensity in light of topology-based considerations. The exciton/polariton strong coupling that has been observed in (iv) naturally incorporates losses due to the coupled system's finite lifetimes. Systems exhibiting this behavior match the definition of the Hamiltonian non-Hermitian, in which the winding numbers $w$, non-Hermitian topological invariant, are the complex eigenenergies of the diagonalized matrix [16,38]. The multilayers that I have studied are framed in open or multilayer (iii) displays $2\pi$ phase modulation (Fig. 4c). The analysis of the loop (iv) (Fig. 4h) display a more clear variation of the phasor amplitude in $\omega t$ coherently with the raise of the losses caused by the coupling of the polariton and the exciton. Moreover, under the condition of a strong coupling, a full $3\pi$ closed domains under a periodic boundary condition imposed by the laser.

related closed loop (Fig. 5a), each lobe corresponds to a winding energy $w$; the main chain in turn crosses at the points where the related winding region is tangential to the main loop, with the trivial region that collapses in a segment.

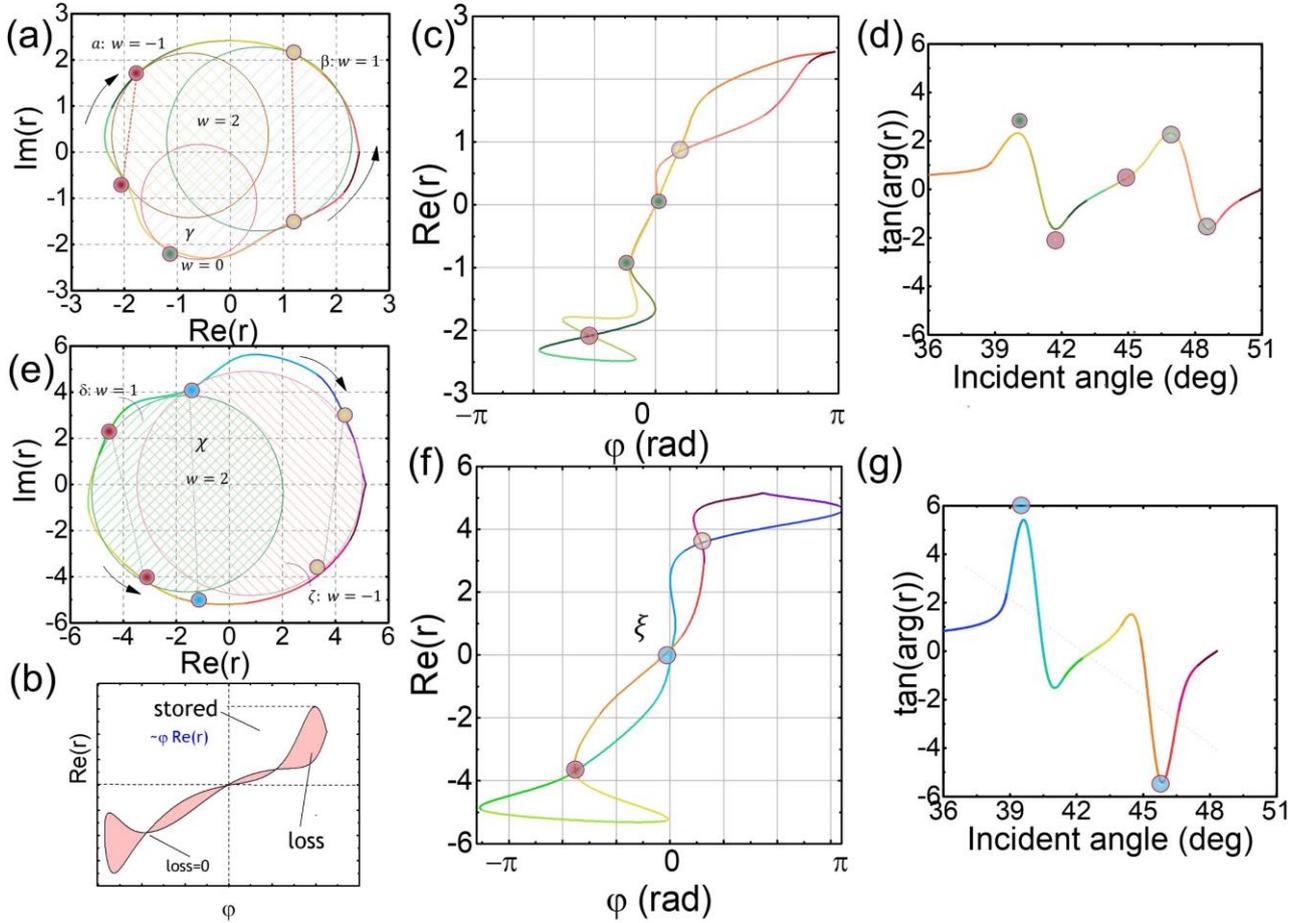

Fig.5. Winding energy. (a,e) representation of $r$ in a complex plane relative to *(iii)* and *(iv)*. The inner dashed circles isolate the regions corresponding to different $w$ with positive or negative signs according to the direction of rotation pointed by the arrows. The circles highlight the tangential points of the inner circles with the main loop. (b) Scheme of $\text{Re}(r) - \varphi$ hysteresis. (c,f) Multiple lobe hysteresis of $\text{Re}(r) - \varphi$. The symbols relate to panels (a,e). (d) $\tan(\arg(r))$ apparent power according to the incident angle.

Besides, because of the Bloch surface waves measured in the asymmetric domains (Supporting Information, Figure S2), the reflectance energy evolves in a loop on the complex plane which can support a topological analysis, rather than collapsing into a real value, as expected in periodic systems[39]. In this context, the multilayers (ii) and (v) are topologically trivial (Figs. 4b, d), the edge modes are negligible, while the cusps at the point where the loop closes confirm the same absolute value shared by several solutions[40]. The configurations (iii) and (iv) exhibit bulk modes at the edges[41][42]; the topology of configuration (iii) in Fig. 5a develops through several closed loops around the origin of the axis, that is the reference point [43] and includes the main and the topologically trivial ($w = 0$) loops. The remaining loops revolve in opposing directions, clockwise ($\alpha: w = -1$) and counterclockwise ($\beta: w = +1$); moreover, inside the region $\gamma$, two linearly independent skin states develop ($w = +2$) and correspond to the complex energy. The variation of the radiative rate, which extracts from the plot in panel (a) the relationship between $\text{Re}(r)$ and the phase $\varphi$, corresponds to by a multiple lobe hysteresis (Fig. 5b). In the hysteresis of the multilayer (iii) (Fig. 5c), built on the color-coded chain of the

All losses introduced into the system alters the energy of (iii) and the lossless points find a correspondence with the apparent power, as depicted by the plot of $\tan(arg(r))$ ($=\frac{Im(r)}{Re(r)}$) versus incident angle (Fig. 5d).

The strong exciton/polariton coupling (multilayer (iv)) develops two closed loops in the complex plane; they progress in opposite directions with reference to the axis origin point (Fig. 5e). The associated winding energy for each complex point in $\delta$ and $\zeta$ regions is $w = \pm 1$, because of the two linearly independent skin states relating to the complex energies. Inside the region $\chi$, the winding energy is $w = +2$, and two linearly independent skin states correspond to the complex energy. The winding energies are polarized at the left ($w = +1$) and right ($w = -1$) edges of the hysteresis (Fig. 5f). The lobes intercross and generate lossless points. It is interesting to observe that the multilayer (iv) crosses a degeneracy point $\xi$ corresponding to the zero-energy state [44], which instantaneously reduces the global energy to zero reaching the point of topological darkness where the reflectance is annihilated [45,46] [47]; this condition corresponds to a range of tunability measured as a $3\pi$ in the phase domain [48]. Another interesting property of $\xi$ is observable in the plot $\tan(arg(r))$ (Fig. 5c). In fact, the response $\tan(arg(r))$ has zero energy magnitude at that point (as proved by the maximal recorded $\tan(arg(r))$), the point corresponds to a phase singularity, which characterizes the reflectivity of (iv) [49]. The zero-state energy that applies to (iv) does not have a counterpart in (iii), which displays a segment in its place. Additionally, (ii) and (v), which in the complex plane exhibit cusps and evolve in a hysteresis loop (Fig.3b and Fig.3e), lack a degeneracy point and do not cross the axis origin (Fig.S4, Supporting information, Part IV). The annihilation of the hysteresis in a single point is then related to the exciton/polariton strong coupling and the identification of $\xi$ figures the topological condition of the strong coupling for the multilayer (iv).

## Experimental

ITO-coated cover slides were utilized as the substrate; to graft the Ag layer atop SiO$_2$ or ITO, a film of (3-mercaptopropyl) trimethoxysilane (Sigma Aldrich) was preliminarily built by chemical vapor deposition in a home-made furnace including vacuum, temperature and pressure control (Table 2, and Supporting information, Part I) [14]. The reflectance in the angular space was measured in a Kretschmann geometry. The tunable continuous-wave laser was collimated (Thorlabs RC12FC-P01) and polarized by a double Glan-Taylor Calcite Polarizer; the noise was lowered with an optical iris diaphragm (Thorlabs, D25SZ). A Si photodiode (λ=960 nm bandwidth) gathers the light signal and a ceramic disk capacitor cuts the noise before delivering the reflected signal to the analyzer [21,22,50]; in addition, the signal was recorded after reaching the steady state. The transfer matrix method was used to model the dispersion of the mode system, which relies on a 4 × 4 matrix for the electromagnetic description of each layer [51]. The properties of the layer were taken from Palik [52] (Ag, ITO, (3-mercaptopropyl) trimethoxysilane) and Schott (NBK7 for the prism, SiO$_2$), while experimental parameters calibrated the model. Table 2 summarizes the material, thickness and refractive index of the layers.

**Table 2. Multilayer layout and layer thickness**

| Layer | Material | Thickness | Refractive index |
|---|---|---|---|
| 1 | ITO | 80-100 nm | 1.83+0.0031*i |
| 2 | SiO$_2$ | 500 mm | 1.52 |
| 3 | (3 mercaptopropyl) trimethoxysilane | 0.1 nm | 1.44 |
| 4 | Ag | 50-70 nm | 0.052+3.9*i |
| 5 | Rhodamine | 5-8 nm | 1.40+0.024*i |

## Conclusions

The reflectance of several multilayer designs, including layers with various functions depending on the selected material, has been studied in this work. An open cavity is generated by the hybridization of resonance modes, which occurs when SiO$_2$ is sandwiched between ITO and Ag (multilayer (iii)). This condition is fundamental for the strong coupling between the polaritons and the excitons when the rhodamine is adsorbed on the surface of the Ag. The exciton/polariton strong coupling is demonstrated by the anti-crossing observed in a dispersion diagram, which in turn enables the multilayer to be categorized as a non-Hermitian system. Based on the winding energy under periodic boundary conditions, a topological investigation of the reflectance of every multilayer here presented has been carried out. The results show that the exciton/polariton strong coupling results in a point of topological darkness, which represents a new paradigm to achieve in the design of multilayers. On the other hand, to produce the rapid phase variation needed in sensing and information technologies, the topological darkness that exists in relation to exciton/polariton strong coupling avoids the restriction on a light amplitude zero, limited to specific systems, and widens the horizons of this field of research.

## Conflicts of interest

There are no conflicts to declare.

## Acknowledgments

The author thanks Dr. Dimitra Antonopoulou, Department of Physical, Mathematical and Engineering Sciences, University of Chester, for the supportive discussions.

## Notes and references

**Symbols**  **Definition**
$E(r)$ electric field of the non-hybridized resonance
$\omega_0$ non-hybridized resonance frequency of the electromagnetic mode
$\varepsilon(r), \Delta\varepsilon(r)$ permittivity distribution, variation
$Im(\varepsilon)$ imaginary component of the permittivity
$\Lambda$ spectral width
$\Gamma$ dissipative loss rate
$\gamma_r$ radiative rate
$\max(r), \min(r)$ resonance maximum, minimum
$C$ constant
$\Delta\omega$ frequency shift
$w$ winding energy
$\varphi$ phase shift
$Im(r), Re(r)$ imaginary, real part of $r$
$Im(E(r)), Re(E(r))$ imaginary, real part of $E(r)$, electric field
$r$ resonance phasor (amplitude of the periodic signal)
$\xi$ zero point energy
$\alpha, \beta, \gamma, \delta, \zeta, \chi$ winding regions
$k_x$ wavenumber
$x, y, z$ spatial coordinates
TE transverse electric

| TM | transverse magnetic |

**Part I**

Ag features have been produced using a two-stage process, with the first phase based on the generation of particles as clusters of nucleation and the second on the growth of the particles.

The chemicals were used without any purification. NaCl (>99%), AgNO$_3$, polyvinylpyrrolidone (PVP, 40000 MW), and ethylene glycol (anhydrous, 99.8%) were purchased from Sigma-Aldrich. Ultrapure water with a resistivity of 18.2 MΩ cm was obtained from an mQ Integral Water Purification System from Merck Millipore. To prevent effects due to the photosensitivity of the AgCl, the synthesis of the salt was performed in a dark environment. A silver nitrate aqueous solution (1 mL, 0.5 M) was mixed with a sodium chloride aqueous solution (1 mL, 1 M). (van de Donk et al. 2019; Zhang and Simone 2019). Following the addition of sodium chloride, flocculation started. The precipitate was separated by the supernatant and washed with fresh and ultrapure water before it was dried. The preparation of the Ag multishape features was performed in a three-neck round-bottom flask at 160 °C and under stirring. The PVP (0.2 mg) was dissolved in 10 mL of ethylene glycol and mixed until reaching a stable temperature; then, the AgCl was added all at once. Inside the sealed flask, the synthesis was left to run for 30 minutes (Simone 2023). To prepare the plasmonic samples, the functional mode (3- mercaptopropyl) trimethoxysilane, which served as the functional molecule, was used to create a chemical bridge between the Ag features and the surface of an ITO/SiO$_2$ substrate. The (3-mercaptopropyl) trimethoxysilane (Sigma-Aldrich) was used to graft the Ag multishape features to create the optical multilayer. It was deposited by chemical vapor deposition (5 µL at 70 °C and 14.7 psia for 30 min) in a homemade furnace that included a vacuum system and a temperature and pressure control. Cover slides with an indium tin oxide ITO-coated side were utilized as the substrate, and the (3-mercaptopropyl) trimethoxysilane was deposited on the SiO$_2$. The substrates were cleaned with a mild stream of toluene or methanol and then dried in a gentle flow of air. The pure Ag feature solution was poured on top of the thiol and allowed to dry for 24 h at room temperature. The substrate was treated with ultrasonication for 30 min in water to remove the multishape features that were not grafted onto the substrate (Figure S1, bottom). Rhodamine G6, a fluorescent dye that may function as a molecular emitter and a radiative antenna, was then adsorbed onto the surface of the Ag layer. When the dye was applied to the surface, it infiltrated through the fissures between the Ag and SiO$_2$ and dried there, as the Ag layer is made of micrometric features aggregated in the real application (Figure 4a). The system becomes more sophisticated after the dye is added, and the optomechanics design includes a spring-mass oscillator made of molecules (Figure S1, top) that are placed on one end mirror of a laser cavity.

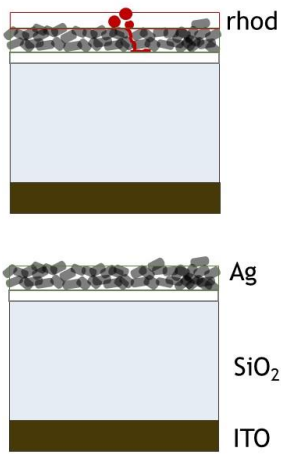

**Figure S1.** Scheme of the arrangement of multilayer (iii) (bottom) and (iv) (top).

**Part II**

For the analysis of the reflectance, the signal was collected by Si photodiode ($\lambda_0$ = 960 nm bandwdth to gather the light, and a ceramic disk capacitor to cut noise) and delivered to an oscilloscope. The signal was analyzed according to the scheme in Figure S1.

**Figure S2.** Left: Scheme of the signal of the voltage versus time; right: phasor of the amplitude.

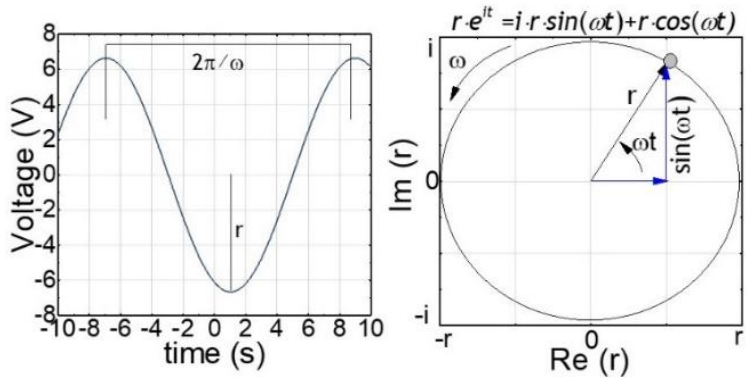

**Part III**

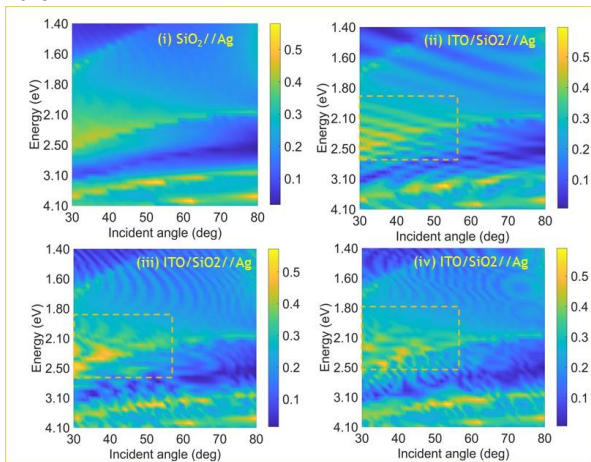

**Figure S3.** Characterization of the plasmonic system. (a) Dispersion for SiO$_2$/ /Ag, ITO/SiO$_2$/ /Ag layout. SiO$_2$ thickness: clockwise 100 mm, 500 mm, and 700 mm. Color bar: reflectance intensity.

**Part IV**

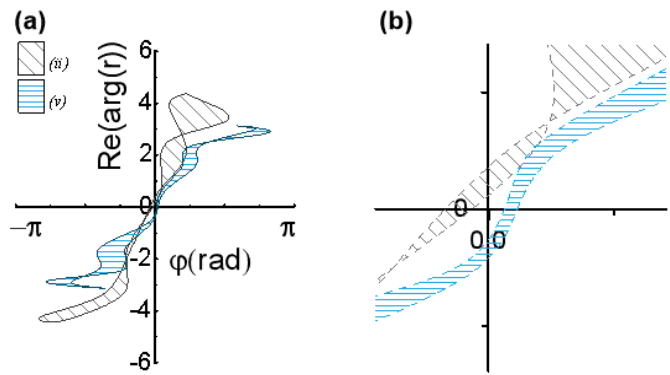

**Figure S4.** (a) Multiple lobe hysteresis of (ii) and (v). (b) The focus on the origin displays the absence of a point of topological darkness.